\documentclass{PoS}
\usepackage{amsmath}
\usepackage{amssymb}
\usepackage{dsfont}
\usepackage{bbm}
\usepackage{subcaption} 
\DeclareMathAlphabet{\mathcal}{OMS}{cmsy}{m}{n}

\title{Towards leading isospin breaking effects in mesonic masses with open boundaries}

\ShortTitle{Towards leading isospin breaking effects in mesonic masses with open boundaries}

\author{\speaker{Andreas~Risch}$^{a}$, Hartmut~Wittig$^{a,b}$\\
        $^{a}$ PRISMA Cluster of Excellence and Institut f{\"u}r Kernphysik, University of Mainz, Germany\\
        $^{b}$ Helmholtz Institute Mainz, University of Mainz, Germany\\
        E-mail: \email{andreas.risch@uni-mainz.de}, \email{hartmut.wittig@uni-mainz.de}}

\abstract{We present an exploratory study of leading isospin breaking effects in mesonic masses using $O(a)$ improved Wilson fermions with open boundaries. Isospin symmetry is explicitly broken by distinct masses and electric charges of the up and down quarks. In order to be able to make use of existing isosymmetric QCD gauge ensembles we apply reweighting techniques. The path integral describing QCD+QED is expanded perturbatively in powers of deviations in the quark masses and the inverse strong coupling as well as the electromagnetic coupling. We have constructed QED$_{\mathrm{L}}$, which we use as a finite volume formulation of QED, for open boundaries. We will also give a first insight into contributions from quark disconnected diagrams.}

\FullConference{The 36th Annual International Symposium on Lattice Field Theory - LATTICE2018\\
		22-28 July, 2018\\
		Michigan State University, East Lansing, Michigan, USA.}

\begin{document}

\section{Introduction}

In recent years, several efforts have been made to investigate isospin breaking effects within the framework of lattice simulations. Isospin is not a perfect symmetry of nature as $m_{\mathrm{u}}\neq m_{\mathrm{d}}$ and $q_{\mathrm{u}}\neq q_{\mathrm{d}}$. Therefore, such investigations have to be performed in a setup that accounts for non-degenerate light quarks including the electromagnetic interaction. Subject of interest are observables that vanish in the isosymmetric limit e.g. the mass splittings of isospin multiplets. Additionally, it is desirable to calculate corrections to high-precision observables traditionally determined in the isosymmetric setup such as the anomalous magnetic moment of the muon.

In this work we adopt the idea of treating isospin breaking effects perturbatively, originally developed by the ROME123 collaboration~\cite{deDivitiis:2011eh, deDivitiis:2013xla}. For a previous account of our effort based on Coordinated Lattice Simulations (CLS) $N_{\mathrm{f}}=2$ QCD ensembles with (anti-)periodic boundaries conditions see~\cite{Risch:2017xxe}. We organise this work as follows: We briefly revisit the setup of QCD+QED simulations based on reweighting and perturbation theory and give a set of Feynman rules for the calculation of correlation functions to leading order isospin breaking. A major part of this project is the formulation of QED$_{\mathrm{L}}$ \cite{Patella:2017fgk, Hayakawa:2008an, Borsanyi:2014jba} on a lattice with open boundary conditions \cite{Luscher:2011kk}. The latter have been employed to generate the CLS $N_{\mathrm{f}}=2+1$ flavour ensembles \cite{Bruno:2014jqa, Bruno:2016plf}, which we use in this work. We propose a hadronic renormalisation scheme to fix the expansion parameters and finally discuss the relevance of QED-connected quark-disconnected diagrams in the mass splittings of pseudo-scalar meson isospin doublets.

\section{Inclusion of perturbative isospin breaking effects by reweighting}

We consider the space of QCD+QED like theories parameterised by $\varepsilon = (m_{\mathrm{u}}, m_{\mathrm{d}},  m_{\mathrm{s}},\beta, e^{2})$ with action $S[U,A,\Psi,\overline{\Psi}] = S_{\mathrm{g}}[U] + S_{\gamma}[A] + S_{\mathrm{q}}[U,A,\Psi,\overline{\Psi}]$, where $S_{\mathrm{q}}[U,A,\Psi,\overline{\Psi}] =  \overline{\Psi}_{\mathbf{a}}{D[U,A]^{\mathbf{a}}}_{\mathbf{b}}\Psi^{\mathbf{b}}$ and $S_{\gamma}[A] = \frac{1}{2}A_{\mathbf{c_{2}}}{\Delta^{\mathbf{c_{2}}}}_{\mathbf{c_{1}}}A^{\mathbf{c_{1}}}$. In position space representation $\mathbf{a},\mathbf{b}\equiv (xfcs)$ and $\mathbf{c}\equiv (x\mu)$ and the metric tensors are unity. For the choice $\varepsilon^{(0)} = (m_{\mathrm{u}}^{(0)}, m_{\mathrm{d}}^{(0)}, m_{\mathrm{s}}^{(0)}, \beta^{(0)}, 0)$ with $m_{\mathrm{u}}^{(0)}= m_{\mathrm{d}}^{(0)}$ we obtain QCD$_{\mathrm{iso}}$ together with a free photon field. QCD+QED can be related to QCD$_{\mathrm{iso}}$ by reweighting~\cite{deDivitiis:2013xla}:
\begin{align}
\langle O[U,A,\Psi,\overline{\Psi}] \rangle &= \frac{\langle R[U] \langle O[U,A,\Psi,\overline{\Psi}] \rangle_{\mathrm{q}\gamma} \rangle_{\mathrm{eff}}^{(0)}}{\langle R[U] \rangle_{\mathrm{eff}}^{(0)}} & R[U] &= \frac{\exp(-S_{\mathrm{g}}[U])Z_{\mathrm{q}\gamma}[U]}{\exp(-S_{\mathrm{g}}^{(0)}[U])Z^{(0)}_{\mathrm{q}}[U]}
\label{eq_expectation_value_by_reweighting}
\end{align}
using the expectation value $\langle O[U,A,\Psi,\overline{\Psi}] \rangle_{\mathrm{q}\gamma} = \frac{1}{Z_{\mathrm{q}\gamma}[U]} \int DA D\Psi D\overline{\Psi} \,\exp\left(-S_{\gamma}[A] - S_{\mathrm{q}}[U,A,\Psi,\overline{\Psi}]\right)$ with $\langle 1 \rangle_{\mathrm{q}\gamma}=1$, the isosymmetric quark determinant $Z^{(0)}_{\mathrm{q}}[U] = \int D\Psi D\overline{\Psi} \,\exp(-S^{(0)}_{\mathrm{q}}[U,\Psi,\overline{\Psi}])$ and the effective expectation value $\langle O[U] \rangle_{\mathrm{eff}}^{(0)} = \frac{1}{Z^{(0)}_{\mathrm{eff}}} \int DU \,\exp(-S^{(0)}_{\mathrm{g}}[U])\,Z^{(0)}_{\mathrm{q}}[U]\,O[U]$ with $\langle 1 \rangle_{\mathrm{eff}}^{(0)}=1$, which is estimated making use of existing QCD$_{\mathrm{iso}}$ gauge configurations. 

In order to evaluate $\left\langle \ldots \right\rangle_{\mathrm{q}\gamma}$ and $Z_{\mathrm{q}\gamma}$ in Eq.~\ref{eq_expectation_value_by_reweighting} we treat isospin breaking effects in the framework of weak coupling perturbation theory~\cite{deDivitiis:2011eh, deDivitiis:2013xla} and expand in terms of $\Delta \varepsilon = \varepsilon - \varepsilon^{(0)}$ around $\varepsilon^{(0)}$. We use the standard graphical representations for the isosymmetric quark and photon propagators:
\begin{align*}
S^{(0)}[U]{{}^{\mathbf{b}}}_{\mathbf{a}} &= (D^{(0)}[U]^{-1}){{}^{\mathbf{b}}}_{\mathbf{a}} = 
\begin{gathered}
\includegraphics[width=5.5em]{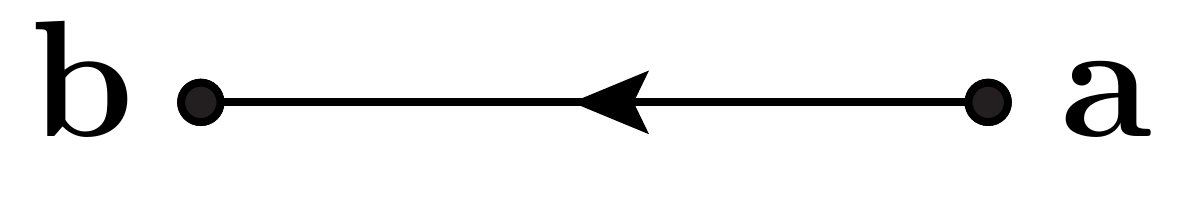}
\end{gathered} & \Sigma{{}^{\mathbf{c_{2}}}}_{\mathbf{c_{1}}} &= (\Delta^{-1}){{}^{\mathbf{c_{2}}}}_{\mathbf{c_{1}}} =
\begin{gathered}
\includegraphics[width=5.5em]{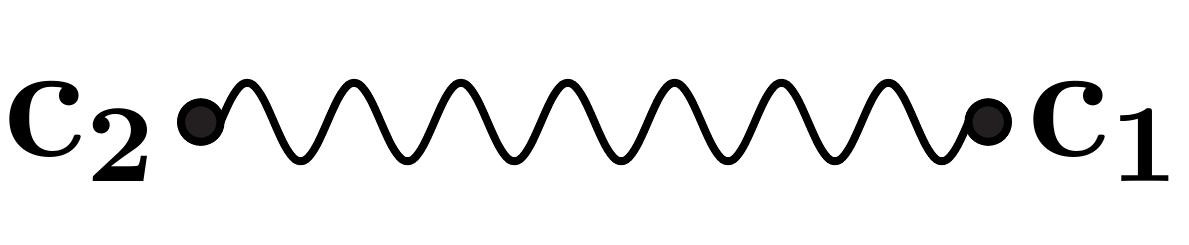}
\end{gathered}
\end{align*}
The Wilson-Dirac-Operator $D[U,A]$ of QCD+QED is expanded around the Wilson-Dirac-Operator \newpage \noindent $D^{(0)}[U]$ of QCD$_{\mathrm{iso}}$. This expansion yields the appropriate vertices for the Feynman rules:
\begin{align*}
D[U,A]{{}^{\mathbf{a}}}_{\mathbf{b}} &= D^{(0)}[U]{{}^{\mathbf{a}}}_{\mathbf{b}} - \sum_{f}\Delta m_{f} 
\begin{gathered}
\includegraphics[width=4.2em]{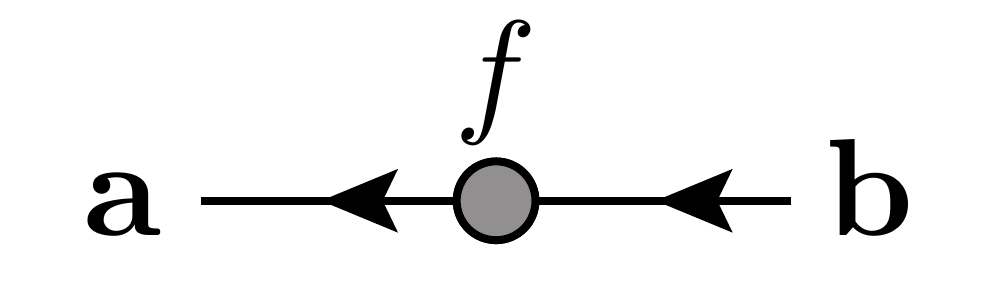}
\end{gathered} - e \begin{gathered}
 \includegraphics[width=4.25em]{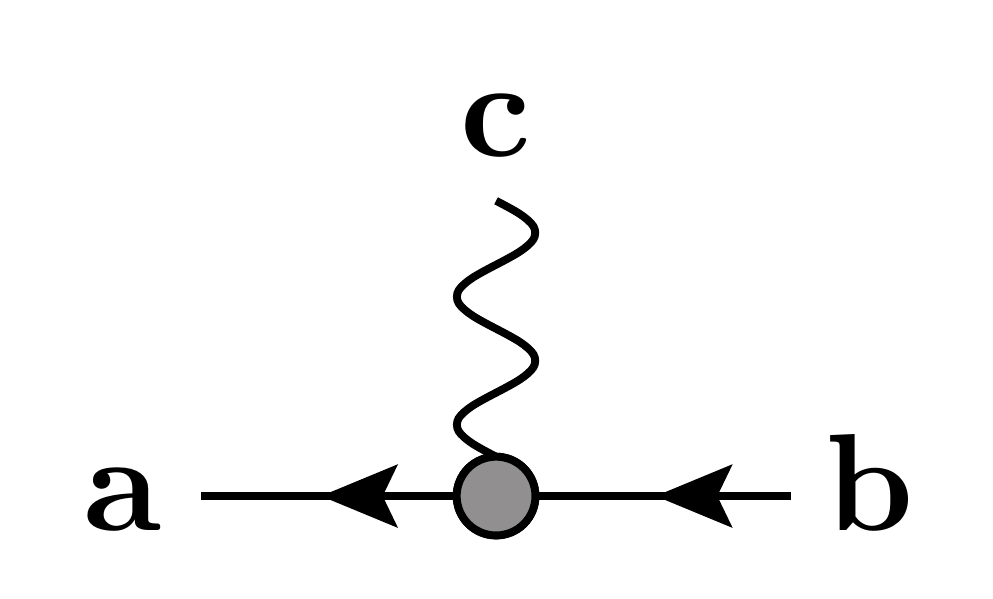}
\end{gathered} A^{\mathbf{c}} - \frac{1}{2} e^{2} \begin{gathered}
\includegraphics[width=4.25em]{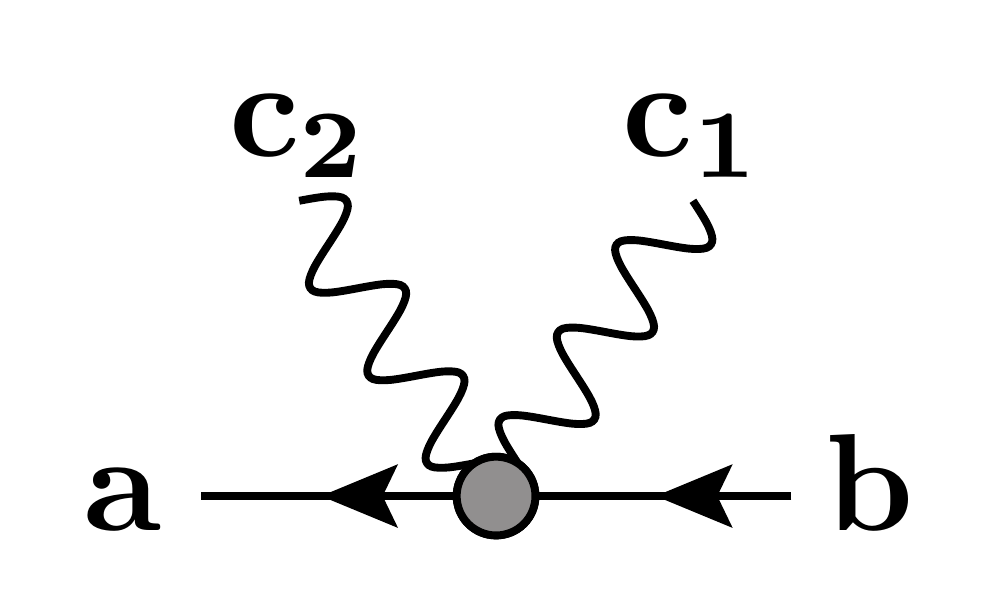}
\end{gathered}A^{\mathbf{c_{2}}}A^{\mathbf{c_{1}}} + O(e^{3})
\end{align*}
In order to expand the reweighting factor $R$ in Eq.~\ref{eq_expectation_value_by_reweighting} we use the standard perturbative expansions of $Z_{\mathrm{q}\gamma}$ and introduce a vertex that accounts for the expansion in $\Delta\beta = \beta-\beta^{(0)}$:
\begin{align*}
\frac{\exp(-S_{\mathrm{g}}[U])}{\exp(-S_{\mathrm{g}}^{(0)}[U])} &= 1 + \Delta\beta \begin{gathered}
\includegraphics[width=2.3333em]{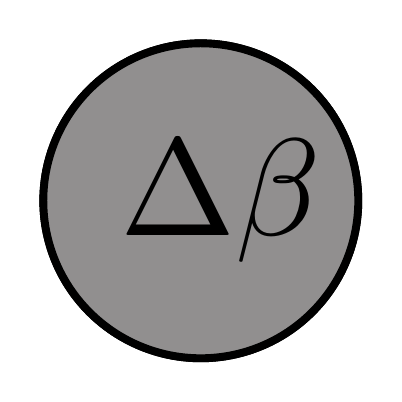}
\end{gathered}
+O(\Delta\beta^{2}) & \begin{gathered}
\includegraphics[width=2.3333em]{vertex_beta.pdf}
\end{gathered}
&= - \frac{1}{\beta^{(0)}}S_{\mathrm{g}}^{(0)}[U],
\end{align*}
where we used the relation $ S_{\mathrm{g}}[U] = \frac{\beta}{\beta^{(0)}} S_{\mathrm{g}}^{(0)}[U]$. Consequently, the reweighting factor is given by
\begin{align*}
\frac{R[U]}{Z^{(0)}_{\gamma}} &= 1 + \sum_{f} \Delta m_{f}
\begin{gathered}
\includegraphics[width=2.5em]{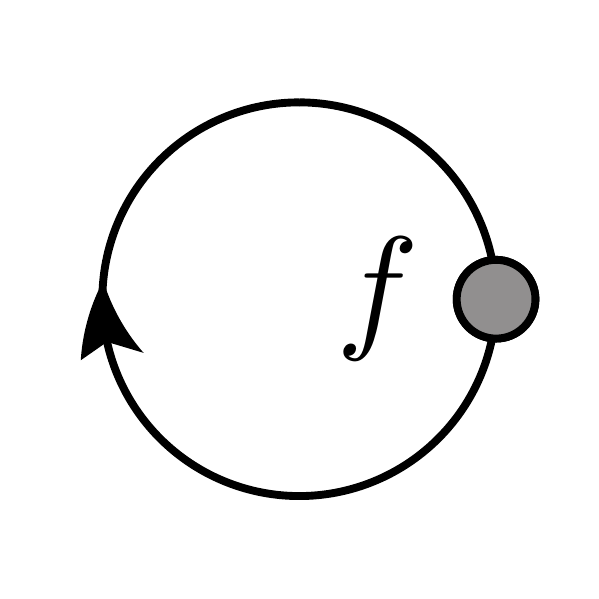}
\end{gathered}
+
\Delta\beta
\begin{gathered}
\includegraphics[width=2.3333em]{vertex_beta.pdf}
\end{gathered}+e^{2}\Big(
\begin{gathered}
\includegraphics[width=4.16666em]{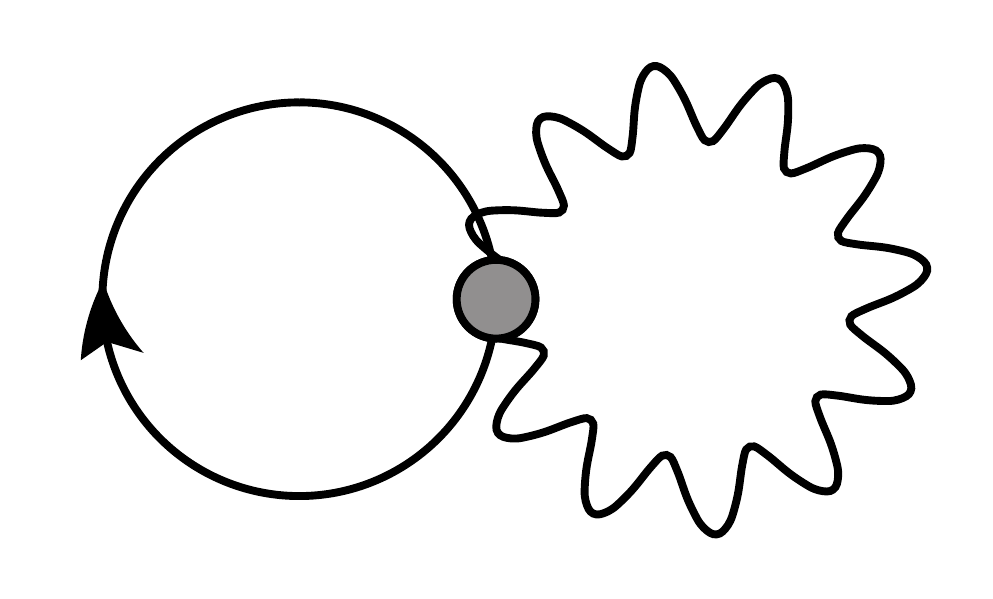}
\end{gathered}
+
\begin{gathered}
\includegraphics[width=2.5em]{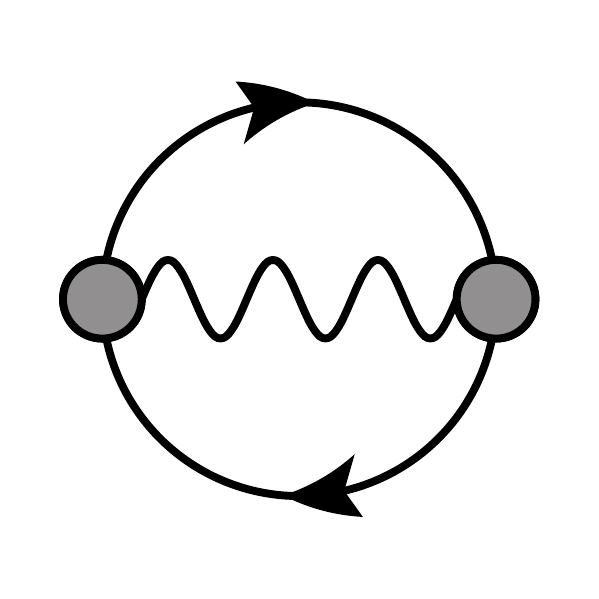}
\end{gathered}
+
\begin{gathered}
\includegraphics[width=5.83333em]{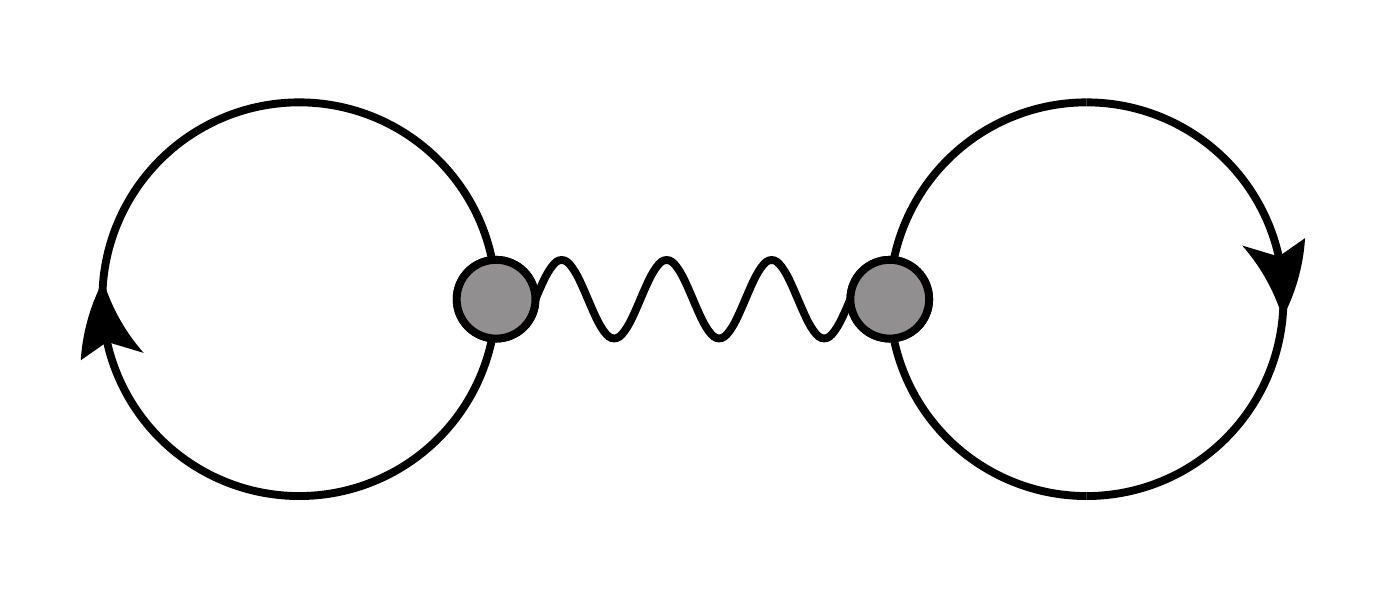}
\end{gathered}
\Big) + O(\Delta\varepsilon^2)
\end{align*}
with $Z^{(0)}_{\gamma}=\int DA \,\exp(-S^{(0)}_{\gamma}[A])$. For each closed fermion loop we have to multiply by factors of $-1$. We also have to divide by the appropriate symmetry factor of the diagram.

\section{QED${}_{\mathrm{L}}$ on open boundaries}

In the following we consider a lattice discretisation $\prod_{\mu=0}^{3}\{0,\ldots,X^{\mu}-1\}$ of $[0,T]\times\mathds{T}^{3}_{L}$ where $X^{0}-1=\frac{T}{a}$ and $X^{\mu}=\frac{L}{a}$ for $\mu=1,2,3$. Consequently, we apply periodic boundary conditions in spatial directions. In the temporal direction we choose open boundary conditions. In analogy to \cite{Luscher:2011kk}, where a compact gauge action was introduced for this lattice topology, we construct a non-compact photon action. We define $S_{\gamma}$ with a gauge fixing term $S_{\gamma,\mathrm{gf}}$, which will be specified later on, as
\begin{align*}
S_{\gamma}[A] &= \frac{1}{4}\sum_{x^{0}=0}^{X^{0}-2}\sum_{\vec{x}}\sum_{\mu=1}^{3}(F^{x0\mu}F^{x0\mu} + F^{x\mu0}F^{x\mu0}) + \frac{1}{4}\sum_{x^{0}=0}^{X^{0}-1}\sum_{\vec{x}}\sum_{\substack{\mu_{1},\mu_{2}=1 \\ \mu_{1}\neq\mu_{2}}}^{3}F^{x\mu_{1}\mu_{2}}F^{x\mu_{1}\mu_{2}} + S_{\gamma,\mathrm{gf}}[A] .
\end{align*}
The absence of $F^{x0\mu}$ and $F^{x\mu 0}$ for $\mu=1,2,3$ in the action at the temporal boundaries leads to the implicit boundary conditions $F^{x0\mu}|_{x^{0}=-1, X^{0}-1} = 0$ i.e. that the electrical field vanishes. This is just the definition of open boundary conditions for the gauge field~\cite{Luscher:2011kk}. Expressing the field strength tensor by the vector potential $F^{\mu_{1}\mu_{2}} = \partial^{\mathrm{f}\mu_{1}}A^{\mu_{2}} - \partial^{\mathrm{f}\mu_{2}}A^{\mu_{1}}$ with $(\partial^{\mathrm{f}\mu}f){}^{x} = f^{x+\mu} - f^{x}$, the condition $F^{x0\mu}|_{x^{0}=-1, X^{0}-1} = 0$ can be implemented by imposing homogeneous Dirichlet boundary conditions on the temporal component $A^{x0}|_{x^{0}=-1, X^{0}-1} = 0$ and homogeneous Neumann boundary conditions on the spatial components $(\partial^{\mathrm{f}0}A^{\mu}){}^{x}|_{x^{0}=-1, X^{0}-1} = 0$ for $\mu=1,2,3$. In fact, this choice is a partial gauge fixing. An appropriate gauge transformation can always be constructed.

In order to eliminate all non-dynamic degrees of the vector potential we redefine the lattice derivatives. Let $\partial^{\mathrm{f}\mu}_{\mathrm{P}}$ and $\partial^{\mathrm{b}\mu}_{\mathrm{P}}$ denote the standard forward (f) and backward (b) derivatives with periodic (P) boundary conditions for $\mu=1,2,3$. For the temporal direction we define derivatives that obey homogeneous Neumann (N) and Dirichlet (D) boundary conditions: $(\partial^{\mathrm{f}0}_{\mathrm{N}}f)^{x} = f^{x+\hat 0} - f^{x}$ for $x^{0}\in\{0,\ldots,X^{0}-2\}$, $(\partial^{\mathrm{b}0}_{\mathrm{D}}f)^{x}=f^{x}$ for $x^{0}=0$, $(\partial^{\mathrm{b}0}_{\mathrm{D}}f)^{x}=f^{x} - f^{x-\hat 0}$ for $x^{0}\in\{1,\ldots,X^{0}-2\}$ and $(\partial^{\mathrm{b}0}_{\mathrm{D}}f)^{x}=-f^{x-\hat 0}$ for $x^{0}=X^{0}-1$. The field strength tensor in terms of the latter derivatives reads
\begin{align*}
F^{\mu_{1} 0} &= - F^{0\mu_{1}} = \partial^{\mathrm{f}\mu_{1}}_{\mathrm{P}}A^{0} - \partial^{\mathrm{f}0}_{\mathrm{N}}A^{\mu_{1}} & F^{\mu_{1}\mu_{2}} &= \partial^{\mathrm{f}\mu_{1}}_{\mathrm{P}}A^{\mu_{2}} - \partial^{\mathrm{f}\mu_{2}}_{\mathrm{P}}A^{\mu_{1}} & \mu_{1},\mu_{2} &= 1,2,3
\end{align*}
We use generalised Coulomb gauge with gauge fixing parameter $\xi$ implying
\begin{align*}
S_{\gamma,\mathrm{gf}}[A] &= \frac{1}{2\xi}\sum_{x^{0}=0}^{X^{0}-1}\sum_{\vec{x}}\sum_{\mu_{1},\mu_{2}=1}^{3}(\partial^{\mathrm{b}\mu_{1}}_{\mathrm{P}}A^{\mu_{1}})^{x}(\partial^{\mathrm{b}\mu_{2}}_{\mathrm{P}}A^{\mu_{2}})^{x}.
\end{align*}
After partial integration, under which $\partial^{\mathrm{f}0}_{\mathrm{N}}$ and $\partial^{\mathrm{b}0}_{\mathrm{D}}$ are transformed into each other, we can read of the photon operator $\Delta$. 

We define basis transformations for open boundary conditions which block-diagonalise $\Delta$. For the spatial directions $\mu=1,2,3$ we define the Fourier transformation $\mathcal{F}{{}^{p^{\mu}}}_{x^{\mu}} = \frac{1}{\sqrt{X^{\mu}}}\exp(-\mathrm{i}p^{\mu}x^{\mu})$ and $(\mathcal{F}^{-1}){{}^{x^{\mu}}}_{p^{\mu}} = \frac{1}{\sqrt{X^{\mu}}}\exp(\mathrm{i}p^{\mu}x^{\mu})$ with $x^{\mu}\in\{0,\ldots,X^{\mu}-1\}$ and $p^{\mu}\in\frac{2\pi}{X^{\mu}}\{0,\ldots,X^{\mu}-1\}$. For the temporal direction we introduce the sine and cosine transformations reading
\begin{align*}
\mathcal{S}{{}^{p^{0}}}_{x^{0}} = (\mathcal{S}^{-1}){{}^{x^{0}}}_{p^{0}} &= \sqrt{\frac{2}{X^{0}}}\sin(p^{0}(x^{0}+1))  & 
\begin{split}
x^{0}&\in\{0,\ldots,X^{0}-2\} \\
p^{0}&\in\frac{\pi}{X^{0}}\{1,\ldots,X^{0}-1\}
\end{split} \\
\mathcal{C}{{}^{p^{0}}}_{x^{0}} = (\mathcal{C}^{-1}){{}^{x^{0}}}_{p^{0}} &=
\begin{cases}
\frac{1}{\sqrt{X^{0}}} & p^{0}=0 \\
\sqrt{\frac{2}{X^{0}}}\cos(p^{0}(x^{0}+\frac{1}{2})) & p^{0}\neq 0
\end{cases} & 
\begin{split}
x^{0}&\in\{0,\ldots,X^{0}-1\} \\
p^{0}&\in\frac{\pi}{X^{0}}\{0,\ldots,X^{0}-1\}.
\end{split}
\end{align*}
$\Delta$ becomes block diagonal by the combined basis change $A^{p0}=\mathcal{S}{{}^{p^{0}}}_{x^{0}}\prod_{\sigma=1}^{3} \mathcal{F}{{}^{p^{\sigma}}}_{x^{\sigma}}A^{x0}$ and $A^{p\mu}=\mathcal{C}{{}^{p^{0}}}_{x^{0}}\prod_{\sigma=1}^{3} \mathcal{F}{{}^{p^{\sigma}}}_{x^{\sigma}}A^{x\mu}$ for $\mu=1,2,3$. Inverting $\Delta$ and taking the limit $\xi \to 0$ we obtain the main result of this section, namely the photon propagator in Coulomb gauge on open boundaries, reading
\begin{align*}
{\Sigma^{p}}_{p} &= \frac{1}{(\sum_{\sigma=1}^{3}p^{\mathrm{f}\sigma}_{\mathrm{P}}p^{\mathrm{b}\sigma}_{\mathrm{P}})(p^{\mathrm{f}0}_{\mathrm{N}}p^{\mathrm{b}0}_{\mathrm{D}}+\sum_{\sigma=1}^{3}p^{\mathrm{f}\sigma}_{\mathrm{P}}p^{\mathrm{b}\sigma}_{\mathrm{P}})} \\
&\hphantom{=}\cdot
\begin{pmatrix}
p^{\mathrm{f}0}_{\mathrm{N}}p^{\mathrm{b}0}_{\mathrm{D}}+\sum_{\sigma=1}^{3}p^{\mathrm{f}\sigma}_{\mathrm{P}}p^{\mathrm{b}\sigma}_{\mathrm{P}} & 0 & 0 & 0 \\
0 & p^{\mathrm{f}2}_{\mathrm{P}}p^{\mathrm{b}2}_{\mathrm{P}}+p^{\mathrm{f}3}_{\mathrm{P}}p^{\mathrm{b}3}_{\mathrm{P}} & -p^{\mathrm{f}1}_{\mathrm{P}}p^{\mathrm{b}2}_{\mathrm{P}} & -p^{\mathrm{f}1}_{\mathrm{P}}p^{\mathrm{b}3}_{\mathrm{P}} \\
0 & -p^{\mathrm{f}2}_{\mathrm{P}}p^{\mathrm{b}1}_{\mathrm{P}} & p^{\mathrm{f}1}_{\mathrm{P}}p^{\mathrm{b}1}_{\mathrm{P}}+p^{\mathrm{f}3}_{\mathrm{P}}p^{\mathrm{b}3}_{\mathrm{P}} & -p^{\mathrm{f}2}_{\mathrm{P}} p^{\mathrm{b}3}_{\mathrm{P}}\\
0 & -p^{\mathrm{f}3}_{\mathrm{P}}p^{\mathrm{b}1}_{\mathrm{P}} & -p^{\mathrm{f}3}_{\mathrm{P}}p^{\mathrm{b}2}_{\mathrm{P}} & p^{\mathrm{f}1}_{\mathrm{P}}p^{\mathrm{b}1}_{\mathrm{P}}+p^{\mathrm{f}2}_{\mathrm{P}}p^{\mathrm{b}2}_{\mathrm{P}}
\end{pmatrix}
\end{align*}
with $p_{\mathrm{D}}^{\mathrm{b}0} = 2\mathrm{i}\sin(\frac{p^{0}}{2})$, $ p_{\mathrm{N}}^{\mathrm{f}0} = -2\mathrm{i}\sin(\frac{p^{0}}{2})$, $p_{\mathrm{P}}^{\mathrm{b}\mu} = -\mathrm{i}(1-\exp(-\mathrm{i}{p^{\mu}}))$ and  $p_{\mathrm{P}}^{\mathrm{f}\mu} = -\mathrm{i}(\exp(\mathrm{i}{p^{\mu}})-1)$.
Finally, as an IR regularisation, we eliminate all spatial zero-modes ${\Sigma^{p}}_{p} = 0$ for $\vec{p}=0$~\cite{Patella:2017fgk, Hayakawa:2008an, Borsanyi:2014jba}.

\section{Definition of a hadronic renormalisation scheme for QCD+QED}

We define the average masses $m_{\pi} = \frac{1}{2}(m_{\pi^{+}}+m_{\pi^{0}})$ and $m_{K} = \frac{1}{2}(m_{K^{+}}+m_{K^{0}})$ as well as the mass splitting $\Delta m_{K} = m_{K^{+}}-m_{K^{0}}$. We suggest to determine the shifts of the bare parameters $\Delta\varepsilon=(a\Delta m_{\mathrm{u}}, a\Delta m_{\mathrm{d}}, a\Delta m_{\mathrm{s}}, \Delta \beta,e^{2})$ by introducing an intermediate scheme where dimensionless quantities are matched between the isosymmetric case, QCD$_{\mathrm{iso}}$, and QCD+QED. The scale $a$ is determined by an additional hadronic mass, $m_{H}$, such that $a = (am_{H})(m_{H}^{\mathrm{phys}})^{-1}$. We match
\begin{align*}
\frac{m_{\pi}}{m_{H}} &= \Big(\frac{m_{\pi}}{m_{H}}\Big)^{(0)} & \frac{m_{K}}{m_{H}} &= \Big(\frac{m_{K}}{m_{H}}\Big)^{(0)}
\end{align*}
on each ensemble. The kaon mass splitting is set to its physical value
\begin{align*}
\frac{\Delta m_{K}}{m_{H}} &= \Big(\frac{\Delta m_{K}}{m_{H}}\Big)^{\mathrm{phys}}.
\end{align*}
Furthermore, we demand the scale to be unchanged i.e. $am_{H} = (am_{H})^{(0)}$. Expanding the masses to first order, $am=(am)^{(0)}+\sum_{l}\Delta\varepsilon_{l} (am)^{(1)}_{l}+O(\Delta\varepsilon^{2})$, this is equivalent to demanding the first order corrections of all the matched quantities to vanish. The electromagnetic coupling does not renormalise at this order and can be fixed to $\Delta\varepsilon_{e^{2}}=e^{2}=4\pi\alpha_{\mathrm{em}}$. We obtain a system of linear equations that determines the full set of expansion parameters $\Delta\varepsilon$:
\begin{align*}
\sum_{l}\Delta\varepsilon_{l} (am_{\pi})^{(1)}_{l} &= 0 & \sum_{l}\Delta\varepsilon_{l} (am_{K})^{(1)}_{l} &= 0 \\
\sum_{l}\Delta\varepsilon_{l} (am_{H})^{(1)}_{l}  &= 0 &
\sum_{l}\Delta\varepsilon_{l} (a\Delta m_{K})^{(1)}_{l} &= \frac{(am_{H})^{(0)}}{m_{H}^{\mathrm{phys}}}\Delta m_{K}^{\mathrm{phys}}.
\end{align*}
In this scheme the summation of zeroth- and first-order contributions is avoided. This is advantageous as first-order contributions can be smaller in magnitude than the statistical errors of the respective zeroth-order contributions.

\section{Contribution of quark-disconnected QED-connected diagrams to meson masses}

In the following, we investigate mass differences and averages in meson isospin multiplets. The expansion of a correlation function reads $C = C^{(0)} + \sum_{l}\Delta\varepsilon_{l}C^{(1)}_{l}+O(\Delta\varepsilon^{2})$. We focus on QED-connected diagrams, which we divide into two classes. The quark-connected contributions read
\begin{align}
C^{(0)}_{\mathrm{con}} &= \Big\langle
\begin{gathered}
\includegraphics[width=6em]{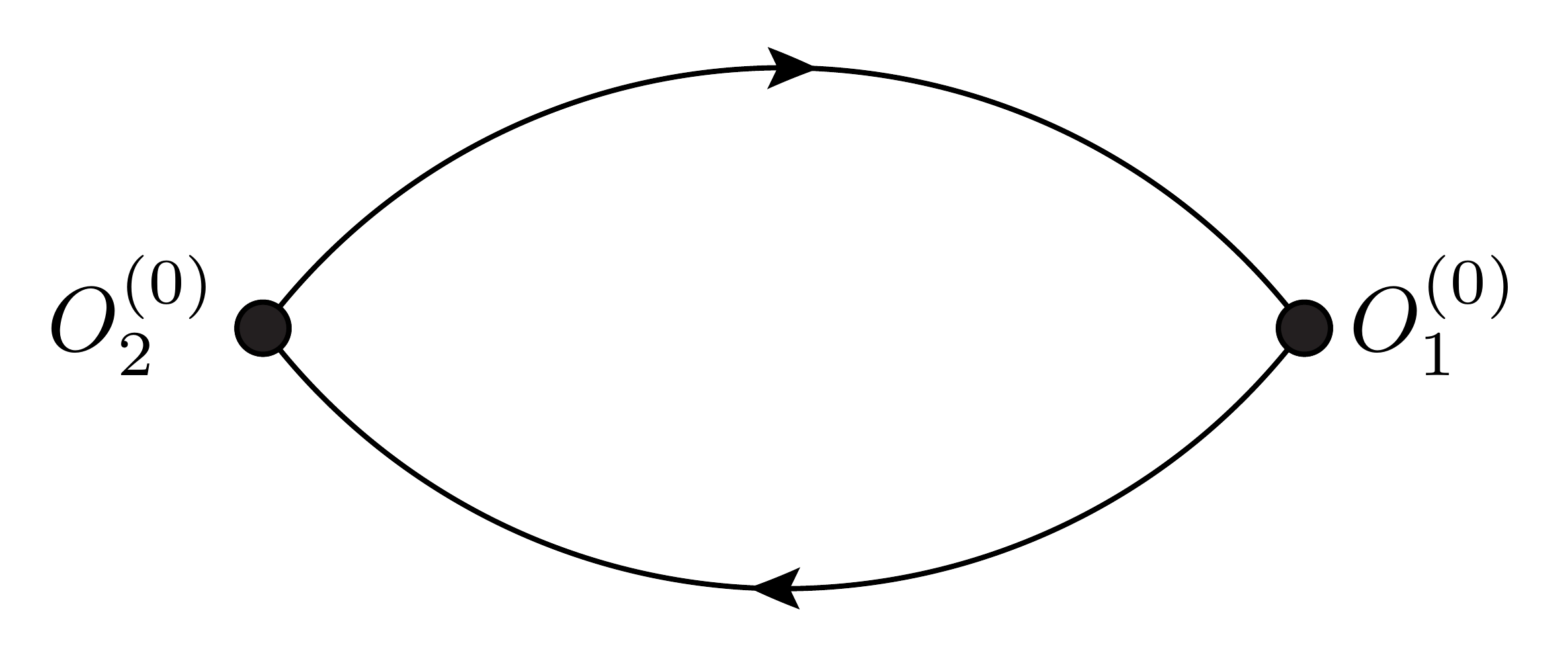}
\end{gathered}
\Big\rangle_{\mathrm{eff}}^{(0)} &
C^{(1)}_{\Delta\mathrm{m}_{f},\mathrm{con,det1}} &= \Big\langle
\begin{gathered}
\includegraphics[width=6em]{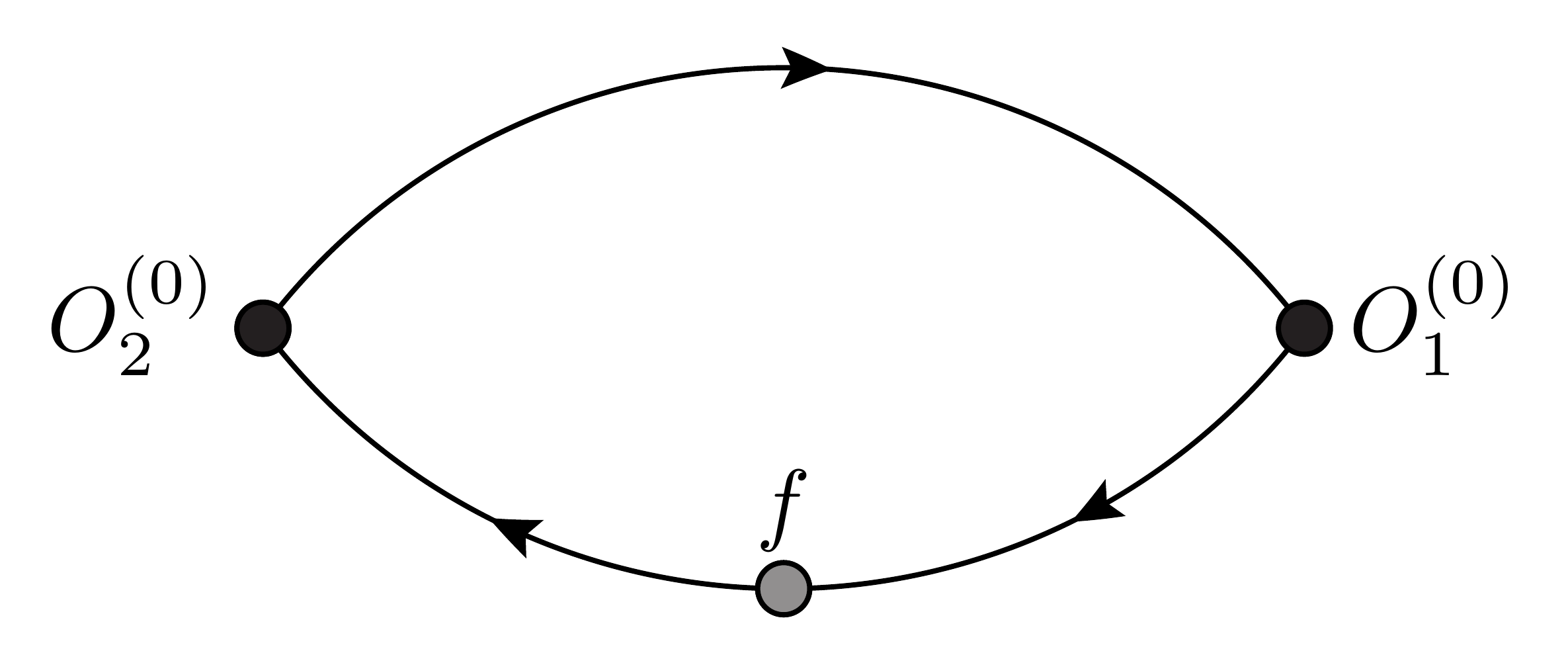}
\end{gathered}
\Big\rangle_{\mathrm{eff}}^{(0)} \nonumber \\
C^{(1)}_{\Delta\beta,\mathrm{con,beta}} &= \Big\langle
\begin{gathered}
\includegraphics[width=6em]{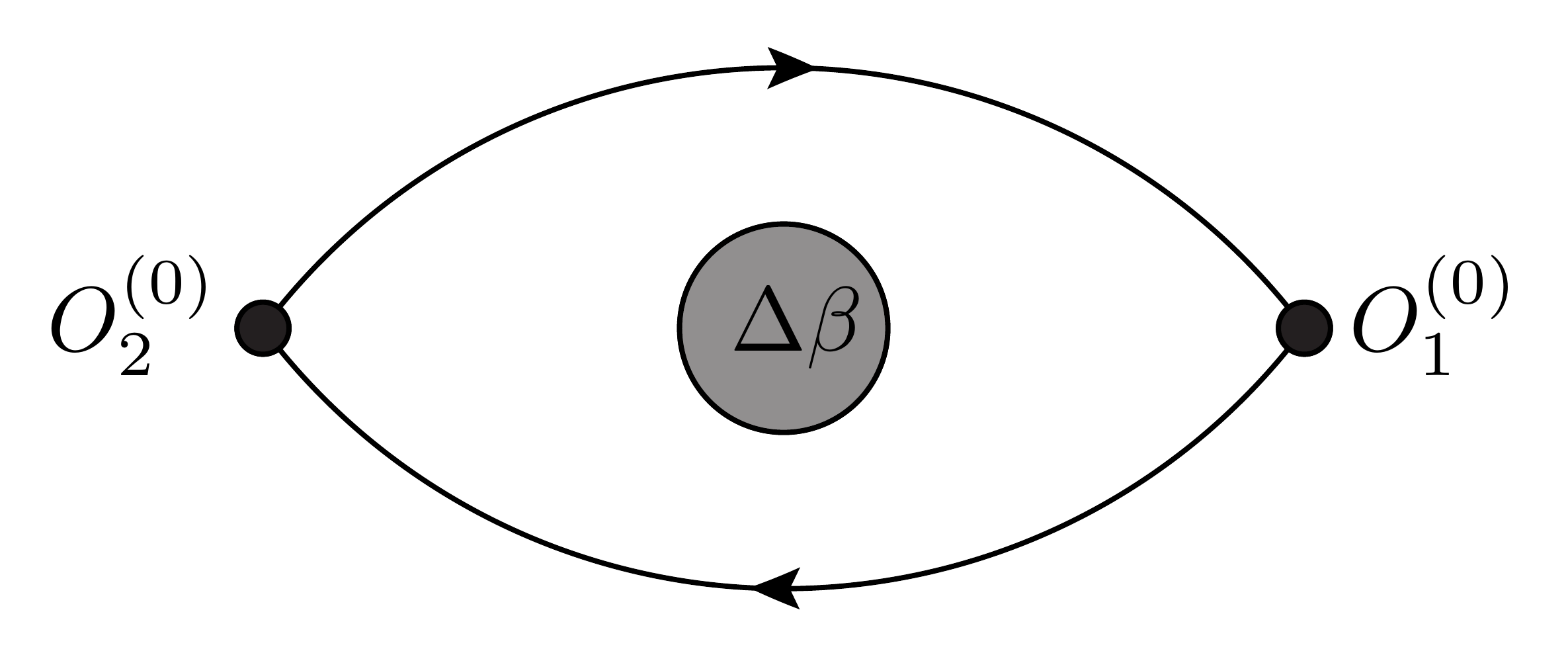}
\end{gathered}
\Big\rangle_{\mathrm{eff}}^{(0)} &
C^{(1)}_{\mathrm{e}^{2},\mathrm{con},\mathrm{exch}} &= \Big\langle
\begin{gathered}
\includegraphics[width=6em]{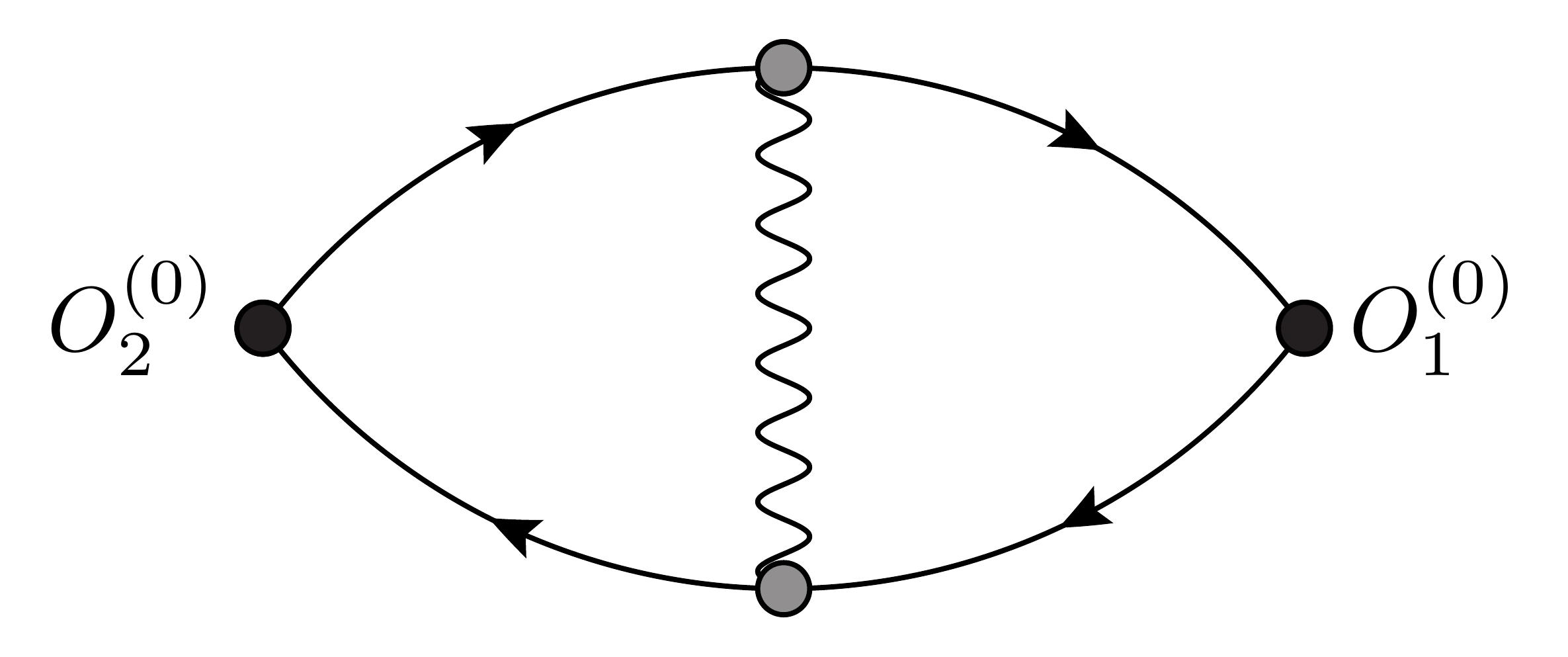}
\end{gathered}
\Big\rangle_{\mathrm{eff}}^{(0)} \nonumber \\
C^{(1)}_{\mathrm{e}^{2},\mathrm{con,self1}} &= \Big\langle
\begin{gathered}
\includegraphics[width=6em]{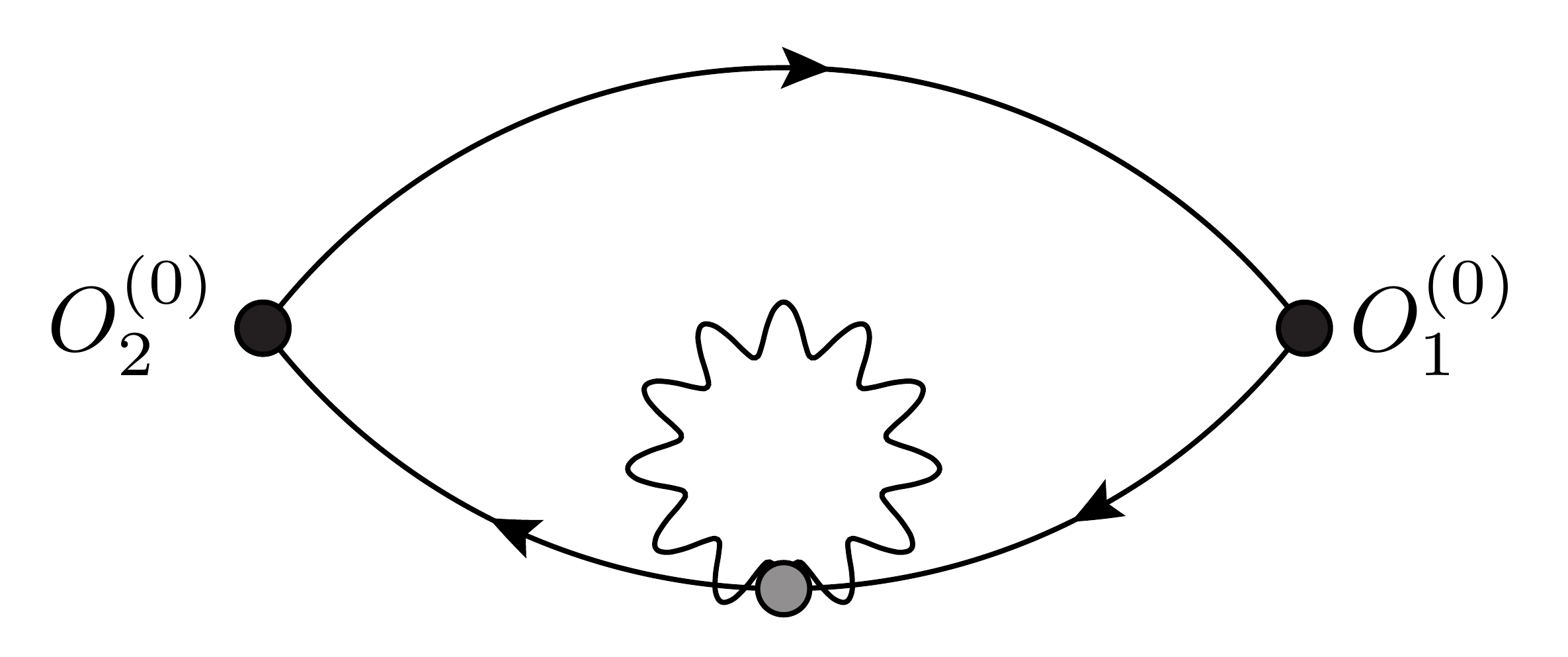}
\end{gathered}
+
\begin{gathered}
\includegraphics[width=6em]{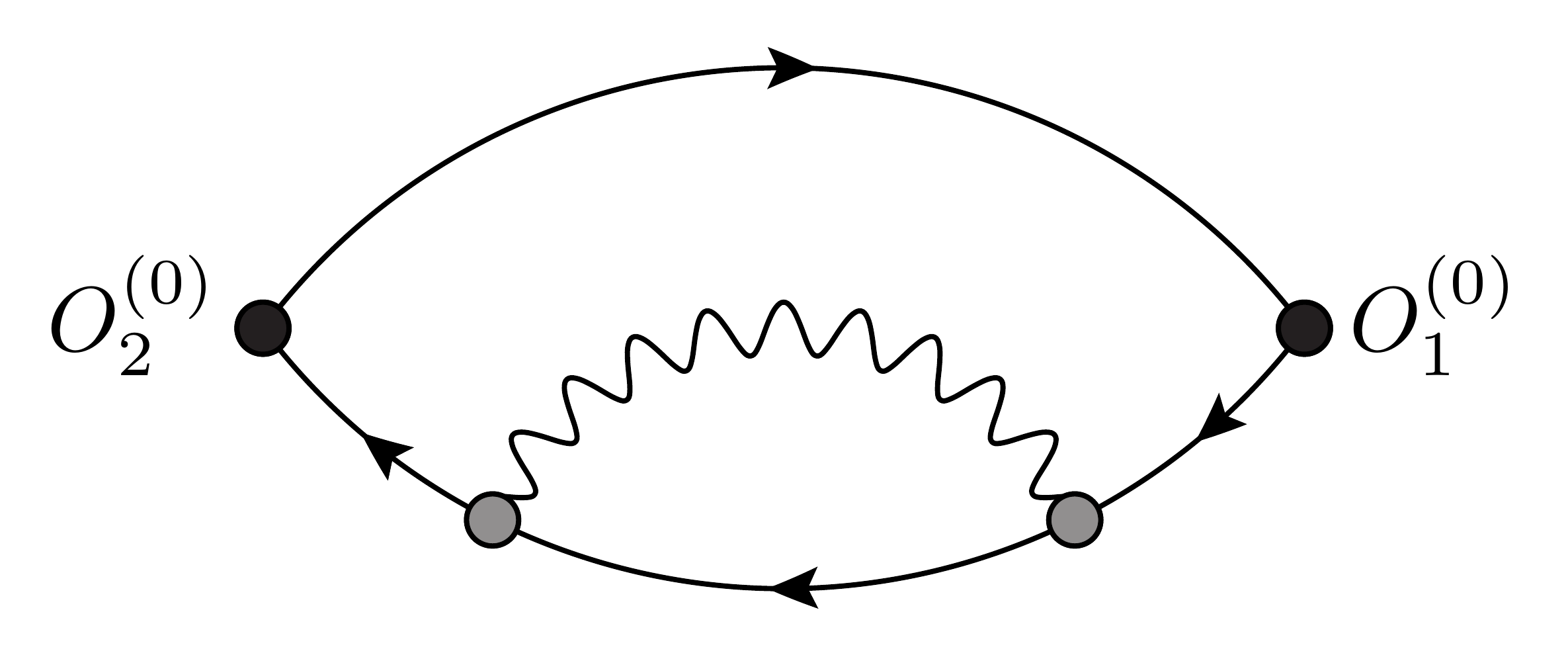}
\end{gathered}
\Big\rangle_{\mathrm{eff}}^{(0)},\hspace{-10em}
\label{eq_qcon_diagrams}
\end{align}
whereas the quark-disconnected QED-connected contributions are given by
\begin{align}
C^{(1)}_{\mathrm{e}^{2},\mathrm{con},\mathrm{vacexch1}} &= \Big\langle
\begin{gathered}
\includegraphics[width=6em]{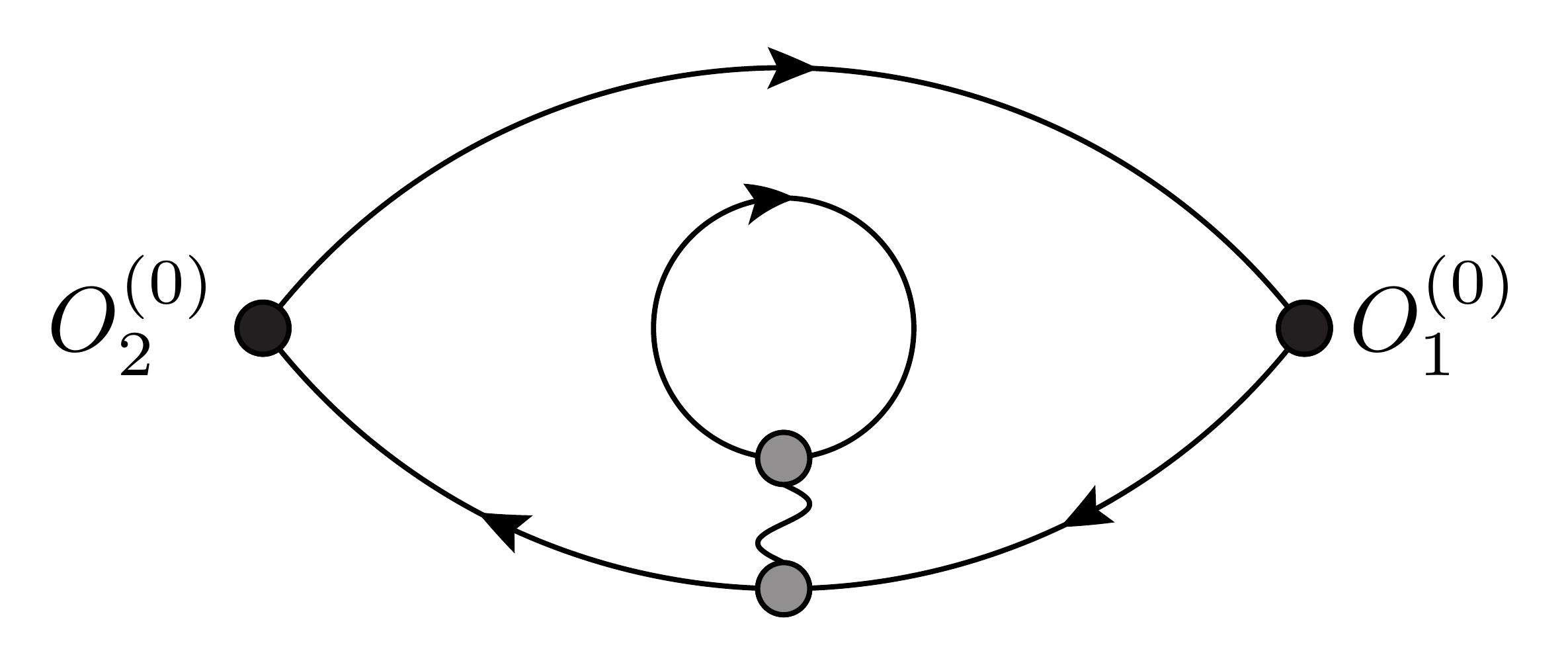}
\end{gathered}
\Big\rangle_{\mathrm{eff}}^{(0)} & 
C^{(1)}_{\mathrm{e}^{2},\mathrm{dis},\mathrm{exch}} &= \Big\langle
\begin{gathered}
\includegraphics[width=6em]{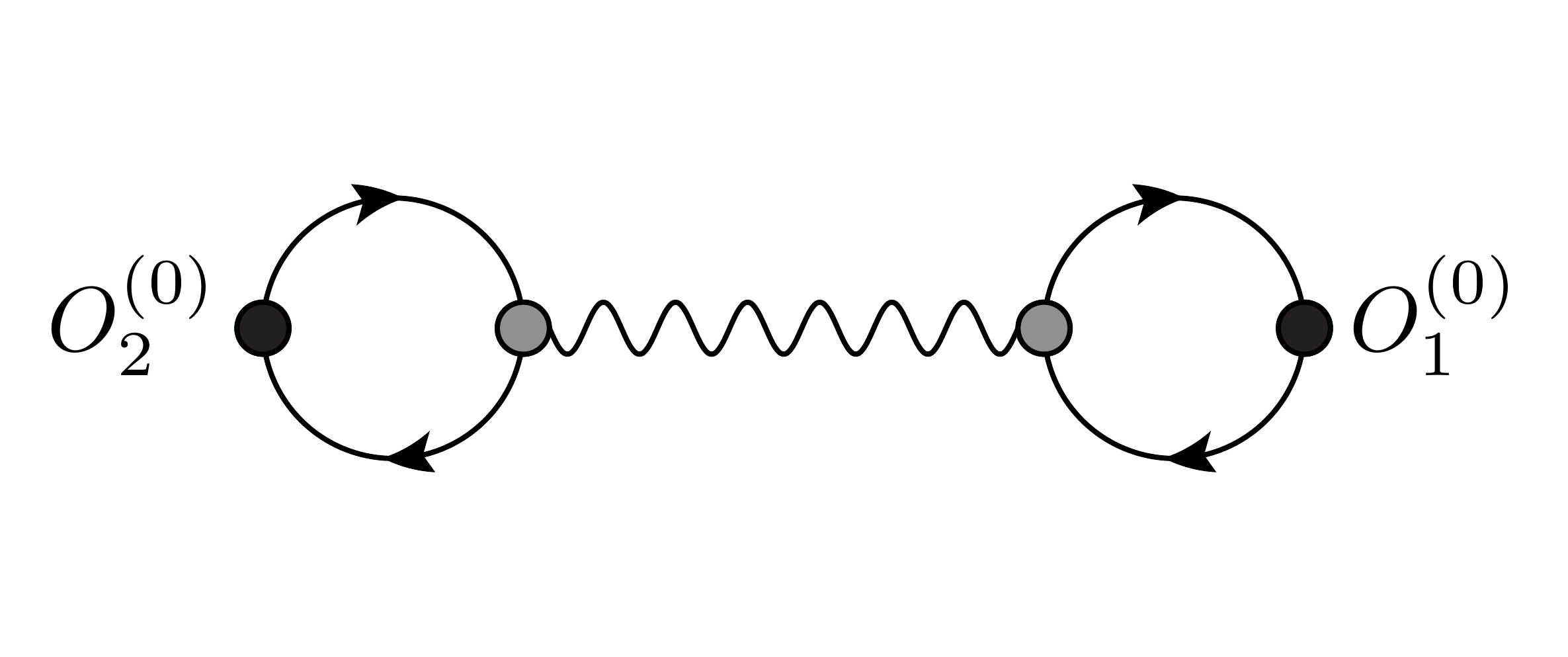}
\end{gathered}
\Big\rangle_{\mathrm{eff}}^{(0)}.
\label{eq_qdis_diagrams}
\end{align}
We omitted the mirrored diagrams $C^{(1)}_{\Delta\mathrm{m}_{f},\mathrm{con,det2}}$, $C^{(1)}_{\mathrm{e}^{2},\mathrm{con,self2}}$ and $C^{(1)}_{\mathrm{e}^{2},\mathrm{con},\mathrm{vacexch2}}$. QED-disconnected diagrams do not contribute to mass splittings for pseudo-scalar isospin meson multiplets and are neglected for mass averages. Expanding the asymptotic behaviour of a correlation function $C(x^{0}) = a\exp(-mx^{0})$ in terms of the parameters $a = a^{(0)}+\sum_{l}\Delta\varepsilon_{l}a^{(1)}_{l}+O(\Delta\varepsilon^{2})$ and $m = m^{(0)}+\sum_{l}\Delta\varepsilon_{l}m^{(1)}_{l}+O(\Delta\varepsilon^{2})$ yields the fit functions
\begin{align}
C^{(0)}(x^{0}) &= a^{(0)}\exp(-m^{(0)}x^{0}) &\frac{C^{(1)}_{l}(x^{0})}{C^{(0)}(x^{0})} &= \frac{a^{(1)}_{l}}{a^{(0)}}-m^{(1)}_{l}x^{0}.
\label{eq_asymptotic_correlation_function}
\end{align}

\begin{table}
  \centering
\begin{tabular}{|l|l|l|l|l|l|l|}
\hline
 & $\mathrm{T}/a \times (\mathrm{L}/a)^3$ & $\beta$ & $a\,[\mathrm{fm}]$ & $m_{\pi}\,[\mathrm{MeV}]$ & $m_{K}\,[\mathrm{MeV}]$ & $m_{\pi}L$ \\ \hline
 H102 & $96 \times 32^{3}$ & $3.4$ & $0.08636(98)(40)$ & $350$ & $440$ & $4.9$ \\ \hline
 H105 & $96 \times 32^{3}$ & $3.4$ & $0.08636(98)(40)$ & $280$ & $460$ & $3.9$ \\ \hline
\end{tabular}
\caption{Parameters of CLS open boundary ensembles with $N_{\mathrm{f}}=2+1$ quark flavours of non-perturbatively O(a) improved Wilson quarks and tree-level improved L\"uscher-Weisz action~\cite{Bruno:2014jqa, Bruno:2016plf}.}
\label{table_lattice_parameters}
\end{table}
\begin{figure}
\centering
\begin{subfigure}[c]{0.49\textwidth}
\includegraphics[width=\textwidth]{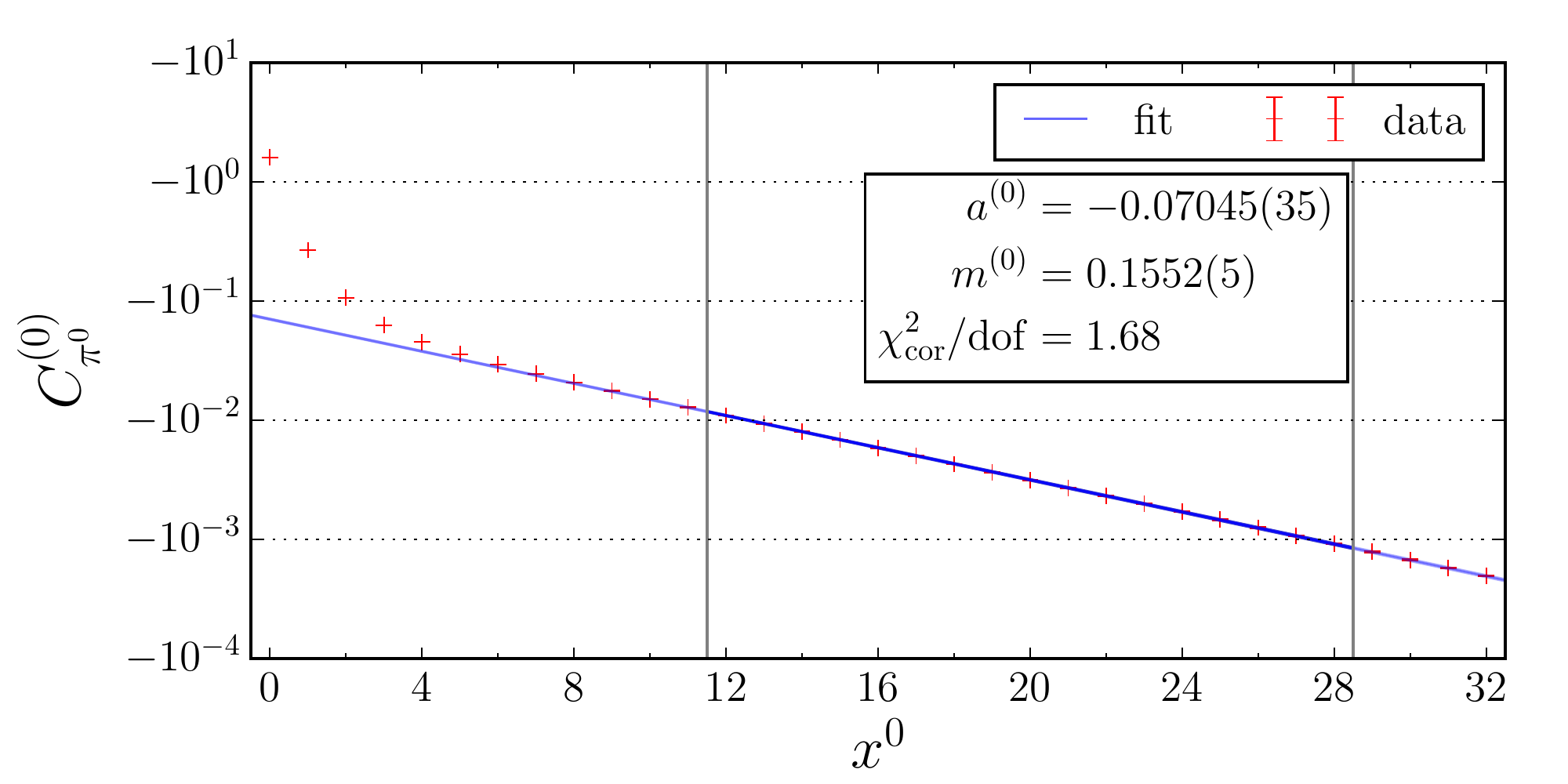}
\subcaption{Connected isosymmetric contribution}
\label{subfig_pion0_con0}
\end{subfigure}
\begin{subfigure}[c]{0.49\textwidth}
\includegraphics[width=\textwidth]{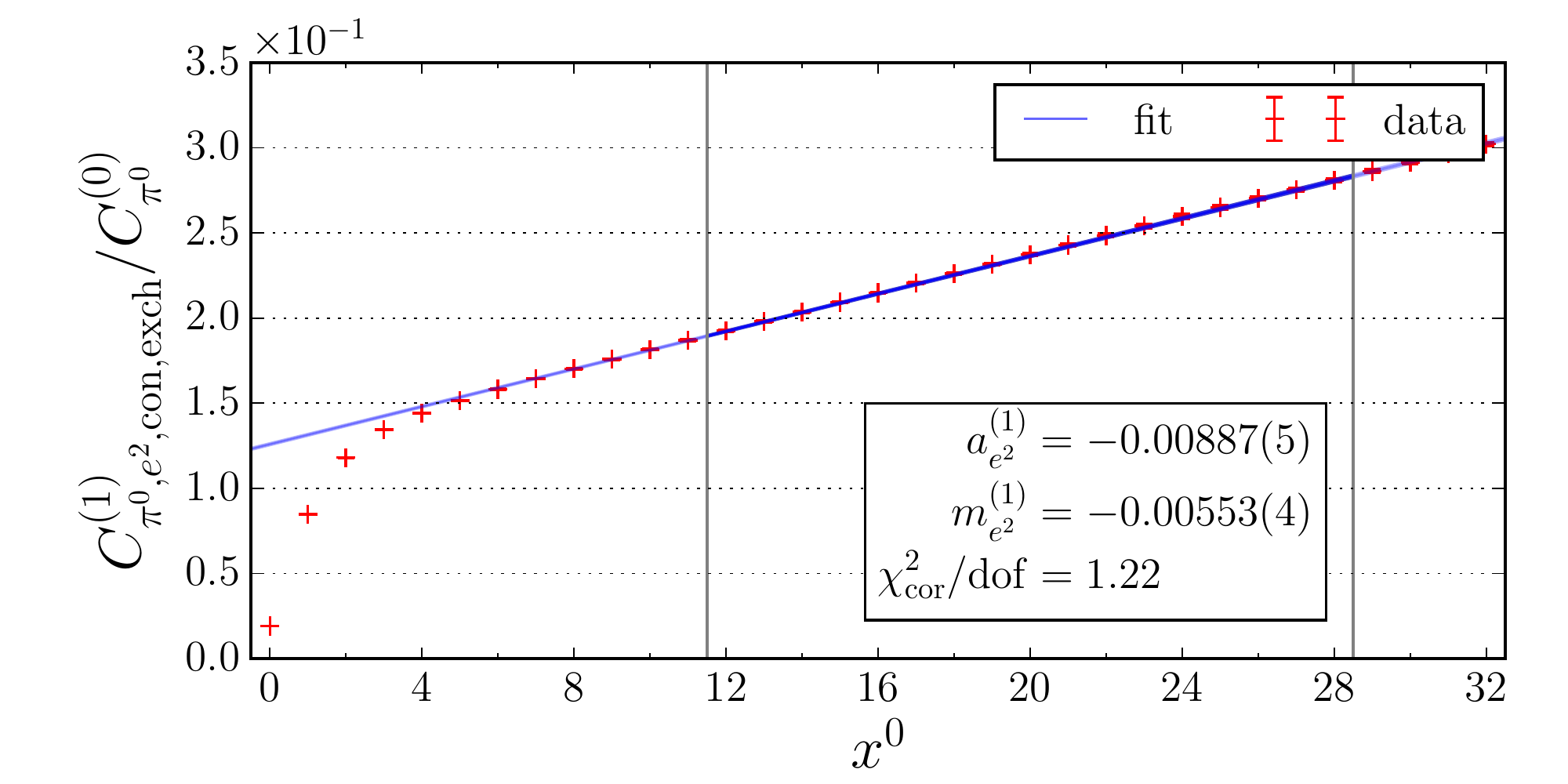}
\subcaption{Connected photon exchange contribution}
\label{subfig_pion0_con1_exch}
\end{subfigure}
\begin{subfigure}[c]{0.49\textwidth}
\includegraphics[width=\textwidth]{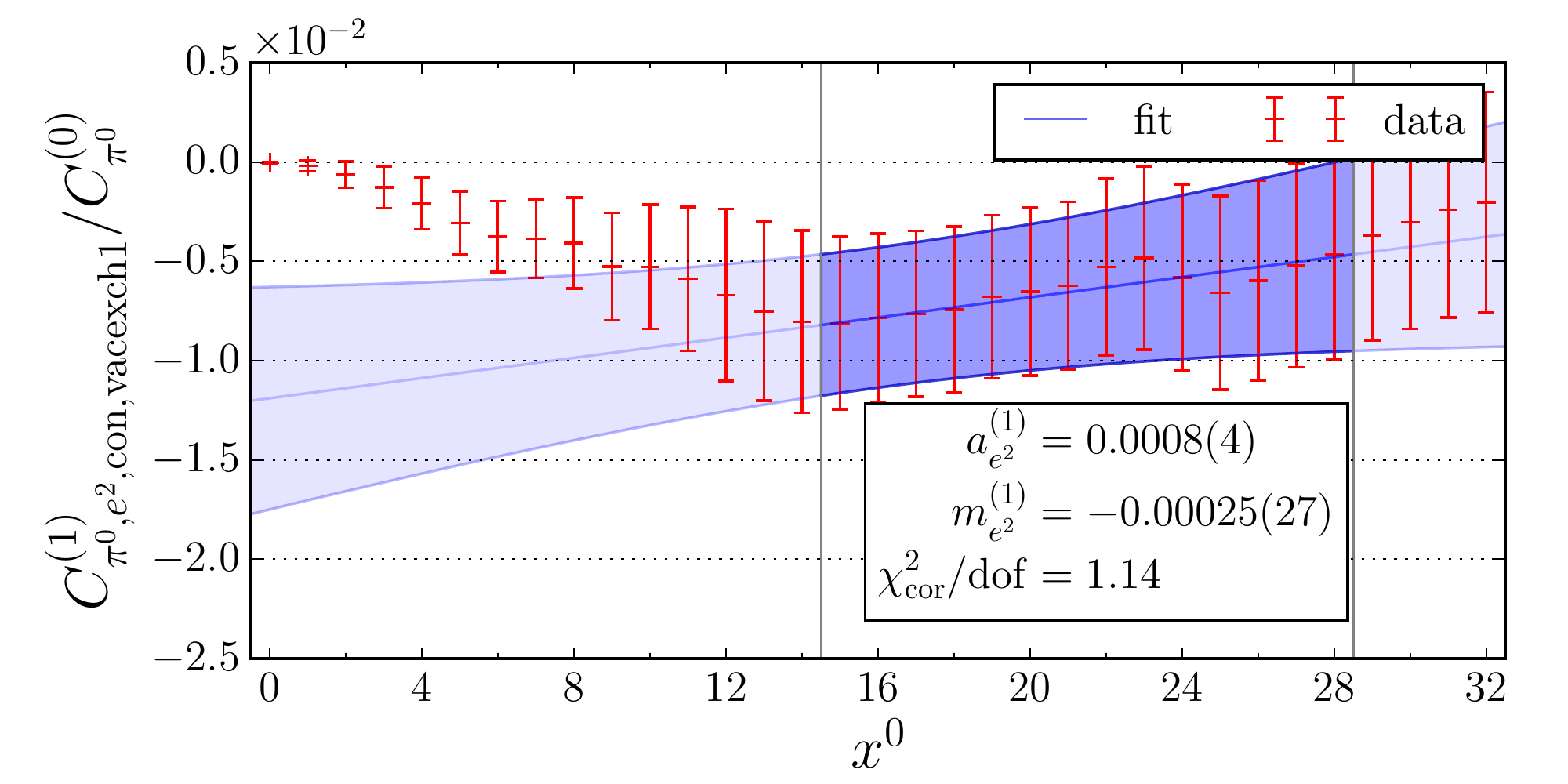}
\subcaption{Connected vacuum exchange contribution}
\label{subfig_pion0_con1_vacexch1}
\end{subfigure}
\begin{subfigure}[c]{0.49\textwidth}
\includegraphics[width=\textwidth]{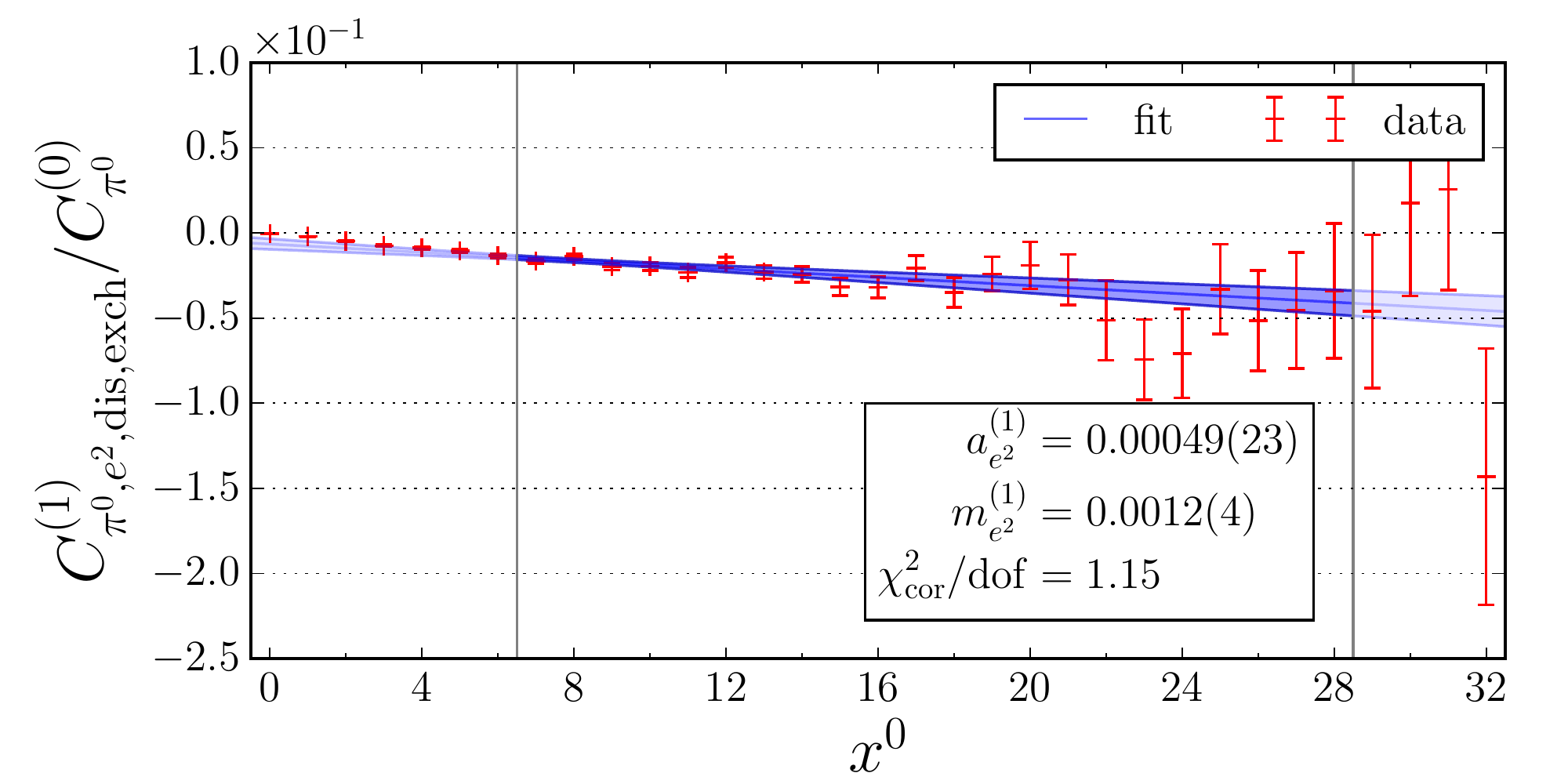}
\subcaption{Disonnected photon exchange contribution}
\label{subfig_pion0_dis1_exch}
\end{subfigure}
\caption[]{Correlation functions for the $\pi^{0}$ meson defined in Eqs.~\ref{eq_qcon_diagrams} and \ref{eq_qdis_diagrams} in lattice units on H102. The fits were performed with respect to Eq.~\ref{eq_asymptotic_correlation_function}.}
\label{fig_pion0}
\end{figure}

The simulation code is based on the OpenQCD~\cite{Luscher:} framework and the QDP++~\cite{Edwards:2004sx} and FFTW3~\cite{FFTW05} libraries. We performed numerical calculations on two gauge ensembles (c.f. Table~\ref{table_lattice_parameters}) on $1000$ configurations, each using $4$ $U(1)$-quark-sources at the meson source, $4\cdot 2$ $Z_{2}$-photon-sources and $4\cdot 8$ $U(1)$-quark-sources at the meson sink for the disconnected loop per configuration.

As an example, Fig.~\ref{fig_pion0} shows contributions for $\pi^{0}$ on H102. The disconnected photon exchange contribution in Fig.~\ref{subfig_pion0_dis1_exch}  contributes at a comparable order of magnitude to the mass as the connected contribution in Fig.~\ref{subfig_pion0_con1_exch} but with opposite sign. The connected vacuum exchange contribution to the mass in Fig.~\ref{subfig_pion0_con1_vacexch1} is of smallest magnitude and equals zero within errors.

\begin{table}
  \centering
\begin{tabular}{|l|l|l|}
\hline
diagrams & H102 & H105 \\ \hline
all & $1.793(149)_{\mathrm{stat}}(48)_{\mathrm{sys}}(22)_{\mathrm{a}}$ & $2.209(154)_{\mathrm{stat}}(52)_{\mathrm{sys}}(27)_{\mathrm{a}}$ \\ \hline
quark-connected & $2.097(21)_{\mathrm{stat}}(52)_{\mathrm{sys}}(26)_{\mathrm{a}}$ & $2.522(58)_{\mathrm{stat}}(24)_{\mathrm{sys}}(31)_{\mathrm{a}}$\\ \hline
\end{tabular}
\caption{$m_{\pi^{+}}-m_{\pi^{0}}$ in $\mathrm{MeV}$ to first order. Fits were performed in the interval $[10,28]$.}
\label{table_pion_splitting}
\end{table}

Making use of Eq.~\ref{eq_asymptotic_correlation_function} the pion mass splitting can be extracted from a fit to $(C^{(1)}_{\pi^{+},\mathrm{e}^{2},\mathrm{con},\mathrm{exch}} - C^{(1)}_{\pi^{0},\mathrm{e}^{2},\mathrm{con},\mathrm{exch}} - C^{(1)}_{\pi^{0},\mathrm{e}^{2},\mathrm{dis},\mathrm{exch}})/C^{(0)}_{\pi,\mathrm{con}}$ defined in Eqs.~\ref{eq_qcon_diagrams} and~\ref{eq_qdis_diagrams}. In the quark-connected approximation $C^{(1)}_{\pi^{0},\mathrm{e}^{2},\mathrm{dis},\mathrm{exch}}$ is neglected. The first-order pion mass splitting is a pure electro-magnetic effect, such that we can use the isosymmetric scale from Table~\ref{table_lattice_parameters} to convert to physical units. The results can be found in Table~\ref{table_pion_splitting}. Systematic fit errors were estimated via the dispersion of the fit results in intervals with boundaries varied by $\pm\{0,1,2\}$. The results differ significantly from the experimentally determined value $m_{\pi^{+}}-m_{\pi^{0}} = 4.5936(5)\,\mathrm{MeV}$~\cite{Patrignani:2016xqp}. This is not unexpected as we simulated at unphysical large pion masses and have neglected the large finite volume corrections caused by the long-range nature of the electromagnetic interaction~\cite{Borsanyi:2014jba, Giusti:2017dmp, Boyle:2017gzv}. However, taking into account that the ensemble H105 is closer to the physical point than H102 with respect to the chosen chiral trajectory, the determined pion mass splittings show the correct behaviour. We also find that for the two chosen lattice ensembles the quark-disconnected contribution is not negligible.

\section{Conclusions and Outlook}

For the investigation of leading order isospin breaking effects using the CLS $N_{\mathrm{f}}=2+1$ open-boundary ensembles we derived an appropriate formulation of QED$_{\mathrm{L}}$ and constructed the photon propagator in Coulomb gauge. We further introduced a renormalisation scheme for matching QCD$_{\mathrm{iso}}$ and QCD+QED, discussed the relevance of quark-disconnected QED-connected contributions to pseudo-scalar meson masses and gave results for the pion mass splitting on two ensembles. We plan to extend our investigations to isospin breaking effects in the hadronic vacuum polarisation for muon $g-2$.
\small
\\
\\
\indent We are grateful to our colleagues within the CLS initiative for sharing ensembles. Our calculations were performed on the HPC Cluster "Clover" at the Helmholtz Institute Mainz and on the HPC Cluster "Mogon II" at the University of Mainz. Andreas Risch is a recipient of a fellowship through GRK Symmetry Breaking (DFG/GRK 1581).
\normalsize

\bibliographystyle{JHEP_mod}
\bibliography{proceedings.bib}

\providecommand{\href}[2]{#2}\begingroup\raggedright\begin{thebibliography}{10}

\bibitem{deDivitiis:2011eh}
G.~M. de~Divitiis et~al.,
  \href{https://doi.org/10.1007/JHEP04(2012)124}{\emph{JHEP} {\bfseries 04}
  (2012) 124} [\href{https://arxiv.org/abs/1110.6294}{{\ttfamily 1110.6294}}].

\bibitem{deDivitiis:2013xla}
G.~M. de~Divitiis et~al.,
  \href{https://doi.org/10.1103/PhysRevD.87.114505}{\emph{Phys. Rev.}
  {\bfseries D87} (2013) 114505}
  [\href{https://arxiv.org/abs/1303.4896}{{\ttfamily 1303.4896}}].

\bibitem{Risch:2017xxe}
A.~Risch and H.~Wittig,
  \href{https://doi.org/10.1051/epjconf/201817514019}{\emph{EPJ Web Conf.}
  {\bfseries 175} (2018) 14019}
  [\href{https://arxiv.org/abs/1710.06801}{{\ttfamily 1710.06801}}].

\bibitem{Patella:2017fgk}
A.~Patella, {\emph{PoS} {\bfseries LATTICE2016} (2017) 020}
  [\href{https://arxiv.org/abs/1702.03857}{{\ttfamily 1702.03857}}].

\bibitem{Hayakawa:2008an}
M.~Hayakawa and S.~Uno, \href{https://doi.org/10.1143/PTP.120.413}{\emph{Prog.
  Theor. Phys.} {\bfseries 120} (2008) 413}
  [\href{https://arxiv.org/abs/0804.2044}{{\ttfamily 0804.2044}}].

\bibitem{Borsanyi:2014jba}
S.~Borsanyi et~al.,
  \href{https://doi.org/10.1126/science.1257050}{\emph{Science} {\bfseries 347}
  (2015) 1452} [\href{https://arxiv.org/abs/1406.4088}{{\ttfamily 1406.4088}}].

\bibitem{Luscher:2011kk}
M.~L{\"u}scher and S.~Schaefer,
  \href{https://doi.org/10.1007/JHEP07(2011)036}{\emph{JHEP} {\bfseries 07}
  (2011) 036} [\href{https://arxiv.org/abs/1105.4749}{{\ttfamily 1105.4749}}].

\bibitem{Bruno:2014jqa}
M.~Bruno et~al., \href{https://doi.org/10.1007/JHEP02(2015)043}{\emph{JHEP}
  {\bfseries 02} (2015) 043} [\href{https://arxiv.org/abs/1411.3982}{{\ttfamily
  1411.3982}}].

\bibitem{Bruno:2016plf}
M.~Bruno, T.~Korzec and S.~Schaefer,
  \href{https://doi.org/10.1103/PhysRevD.95.074504}{\emph{Phys. Rev.}
  {\bfseries D95} (2017) 074504}
  [\href{https://arxiv.org/abs/1608.08900}{{\ttfamily 1608.08900}}].

\bibitem{Luscher:}
M.~L{\"u}scher, {\emph{http://cern.ch/luscher/openQCD/} (2014) }.

\bibitem{Edwards:2004sx}
R.~G. Edwards and B.~Joo,
  \href{https://doi.org/10.1016/j.nuclphysbps.2004.11.254}{\emph{Nucl. Phys.
  Proc. Suppl.} {\bfseries 140} (2005) 832}
  [\href{https://arxiv.org/abs/hep-lat/0409003}{{\ttfamily hep-lat/0409003}}].

\bibitem{FFTW05}
M.~Frigo and S.~G. Johnson, {\emph{Proceedings of the IEEE} {\bfseries 93}
  (2005) 216}.

\bibitem{Patrignani:2016xqp}
C.~Patrignani et~al.,
  \href{https://doi.org/10.1088/1674-1137/40/10/100001}{\emph{Chin. Phys.}
  {\bfseries C40} (2016) 100001}.

\bibitem{Giusti:2017dmp}
D.~Giusti et~al., \href{https://doi.org/10.1103/PhysRevD.95.114504}{\emph{Phys.
  Rev.} {\bfseries D95} (2017) 114504}
  [\href{https://arxiv.org/abs/1704.06561}{{\ttfamily 1704.06561}}].

\bibitem{Boyle:2017gzv}
P.~Boyle et~al., \href{https://doi.org/10.1007/JHEP09(2017)153}{\emph{JHEP}
  {\bfseries 09} (2017) 153}
  [\href{https://arxiv.org/abs/1706.05293}{{\ttfamily 1706.05293}}].

\end{thebibliography}\endgroup

\end{document}